# Selective C−C Coupling by Spatially Confined Dimeric Metal Centers

Yanyan Zhao, Si Zhou[*], Jijun Zhao

**SUMMARY**

Direct conversion of carbon dioxide ($CO_2$) to high-energy fuels and high-value chemicals is a fascinating sustainable strategy. For most of the current electrocatalysts for $CO_2$ reduction, however, multi-carbon products are inhibited by large overpotentials and low selectivity. For practical applications, there remains a big gap of knowledge in proper manipulation of the C−C coupling process. Herein, we exploit dispersed $3d$ transition metal dimers as spatially confined dual reaction centers for selective reduction of $CO_2$ to liquid fuels. Various nitrogenated holey carbon monolayers are shown to be promising templates to stabilize these metal dimers and dictate their electronic structures, allowing precise control of the catalytic activity and product selectivity. By comprehensive first-principles calculations, we screen the suitable transition metal dimers that universally have high activity for ethanol ($C_2H_5OH$). Furthermore, remarkable selectivity for $C_2H_5OH$ against other $C_1$ and $C_2$ products is found for $Fe_2$ dimer anchored on $C_2N$ monolayer. The correlation between the activity and $d$ band center of the supported metal dimer as well as the role of electronic coupling between the metal dimer and the carbon substrates are thoroughly elucidated.

Key Laboratory of Materials Modification by Laser, Ion and Electron Beams (Dalian University of Technology), Ministry of Education, Dalian 116024, China
[*] Correspondence: sizhou@dlut.edu.cn (Si Zhou)

**INTRODUCTION**

Production of liquid fuels by catalytic convertion of $CO_2$, the main greenhouse gas and meanwhile an abundant carbon feedstock, has been regarded as an appealing strategy to solve both energy and environmental crises, albeit facing great challenges (Birdja et al., 2019; Jia et al., 2019; Amal et al., 2017). Copper-based materials have been widely adopted as catalysts for electro-reduction of $CO_2$ to multi-carbon ($C_2$ or $C_{2+}$) products (Zheng et al., 2019). Although fairly good activity can be achieved by modification or morphology engineering of copper, such as sculpturing it into nanoparticles or nanocubes, doping or alloying, and making oxide-derived copper, the selectivity and efficiency of most copper-based electrocatalysts remain unsatisfactory for commercialization of the $CO_2$ conversion technique to high-energy fuels and high-value chemicals (Kim et al., 2017; Zhou et al., 2018; Gao et al., 2019; Wang et al., 2018).

Recenlty, transition metal atoms dispersed on nitrogen-doped porous carbon nanomaterials emerge as a promising category of electrocatalysts for $CO_2$ reduction, which have maximum atomic efficiency, high electrical conductivity and good durability, and can be facilely synthesized in the laboratory (Wang et al., 2019; Bayatsarmadi et al., 2017; Chen et al., 2019; Cheng et al., 2018). The transition metal atoms are usually anchored in the pores of the carbon matrix and coordinated with the nitrogen atoms, exhibiting unique electronic states and acting as isolated reaction centers for $CO_2$ reduction. Remarkable activity and selectivity toward carbon monoxide (CO) has been observed for various dispersed transition metal atoms (Fe, Co, Ni, Mn, and Cu) on N-doped graphene, carbon nanosheets or nanospheres, with selectivity up to 97% and Faradaic efficiency above 80% (Jiang et al., 2018; Zhang et al., 2018; Yang et al., 2018; Wang et al., 2019; Ren and Zhao, 2019). First-principles calculations show that the activity highly depends on the type of metal atoms, which provide different binding strengthes with the reaction intermediates (Ju et al., 2017). The single metal sites also have an advantage of suppressing the competing hydrogen evolution reaction (HER), due to the unique adsorption configuration of H* species compared to those on the transition metal surfaces (Bagger et al., 2017).

Furthermore, homonuclear and heteronuclear dimers of transition metal immobilized in carbon based nanostructures, such as $Fe_2$ and Fe-Co on nitrogenated graphitic carbon materials, Fe-Ni on N-doped graphene, and Pt-Ru on g-$C_3N_4$, have been synthesized in the laboratory (Ye et al., 2019; Wang et al., 2017; Wang et al., 2018; Zhou et al., 2019). This opens up the windows for a broader range of chemical processes that require dual reaction centers either with enhanced activity or carrying different functionalities simultaneously. For instance, Ren et al. fabricated diatomic Fe-Ni sites embedded in nitrogenated carbon (Ren et al., 2019). By taking advantage of the strong binding capability of Fe with $CO_2$ molecule and the weak adsorption of CO on Ni, they achieved impressively high selectivity of 99% for CO and Faradaic efficiency above 90% over a wide potential range from −0.5 to −0.9 V, reaching 98% at −0.7 V vs. reversible hydrogen electrode (RHE). On the theoretical side, a $Cu_2$ dimer supported on the $C_2N$ monolayer was predicted to have high selectivity for methane ($CH_4$), while dimerization of two CO species leading to the formation of ethene ($C_2H_4$) is possible with an energy cost of 0.76 eV (Zhao et al., 2018). Other $C_1$ products have also been proposed for elementary and mixed metal dimers on various N-coordinated carbon nanostructures, including defective graphene and phthalocyanine (Li et al., 2015; He et al., 2018; Shen et al., 2017).

Intuitively, two adjacent metal atoms that are spatially confined in a hole of N-doped carbon materials can provide unique active sites, which not only enable the simultaneous fixation of two $CO_2$ molecules, but also sterically limit the reaction pathways that may be beneficial for C−C coupling toward $C_2$ or $C_{2+}$ products. Moreover, various combinations of metal dimers and carbon substrates give high degrees of freedom for modulating the catalytic performance. However, the atomistic mechanism and composition recipe of such heterogeneous catalysts remain largely unknown, which impede their rational design and experimental synthesis for practical uses.

Here we exploit 3*d* transition metal dimers immobilized on various nitrogenated holey carbon sheets for selective reduction of $CO_2$ to $C_2$ products. By systematic first-principles calculations, the detailed C−C coupling mechanism on the spatially confined dual metal centers has been elucidated for the first time. The suitable transition metal elements and carbon substrates that lead to high activity and selectivity for ethanol ($C_2H_5OH$) and $C_2H_4$ are screened, and the underlying electronic structure-activity relationship is unveiled. These theoretical explorations illuminate important clues for precisely engineering the dispersed metal catalysts on porous carbon nanomaterials for direct conversion of greenhouse gas to multi-carbon hydrocarbons and oxygenates.

**RESULTS and DISCUSSION**

In the laboratory, N-doped graphitic carbon materials with controllable doping contents (up to 16.7% of N content) and atomic geometries can be achieved via either direct synthesis or post treatment (Xue et al., 2012; Xu et al., 2018; Qu et al., 2010). Here we focused on pyridine N dopants in graphene, which are the main doping species at high N contents and are usually associated with the vacancies or pores of the carbon basal plane (Sheng et al., 2011; Sarau et al., 2017). As displayed in Figure 1, we considered a series of N-doped holey graphene monolayers, comprising C vacancies of various sizes (denoted as $V_n$, $n$ = 2, 3, 4, 6) with the edges coordinated with different numbers of N atoms (denoted as $m$N, $m$ = 4, 5, 6). Specifically, 4N-$V_2$, 5N-$V_3$ and 6N-$V_4$ systems can be viewed as four, five and six N atoms decorating the edges of di-vacancy, tri-vacancy and tetra-vacancy in graphene, respectively, all of which have been commonly observed in experiment (Lin et al., 2015; He et al., 2014; Wang et al., 2018). Note that, the $V_6$ pore in graphene is a favorable defect according to transmission electron microscopy experiment (Robertson et al., 2015), and our previous calculation showed that N-doped $V_6$ (namely 6N-$V_6$) has extraordinary thermodynamic stability (Luo et al., 2013). Besides the N-doped graphitic sheets, we also considered the synthetic carbon nitride monolayers, including g-$C_3N_4$ and $C_2N$ (Zhao et al., 2014; Mahmood et al., 2015). All these porous N-coordinated carbon sheets have formation energies (defined by Equation S1 in Supplemental Information) in the range of 0.16 ~ 0.21 eV/Å, while the N-free $V_6$ is higher in energy by over 0.46 eV/Å than the others (Table 1). These nitrogenated 2D holey carbon materials are ideal templates to stabilize and disperse metal atoms or small clusters. Indeed, isolated $Fe_2$, Fe-Ni, and Fe-Co dimers embedded in 6N-$V_4$(a), as well as $Fe_2$ and Pt-Ru dimers anchored on g-$C_3N_4$ have already been realized in experiment (Ye et al., 2019; Wang et al., 2017; Zhou et al., 2019; Ren et al., 2019; Tian et al., 2018).

To evaluate the capability of various supported metal dimers for $CO_2$ reduction toward $C_2$ products, we first explored the atomic structures, electronic and adsorption properties of dimeric $3d$ transition metal clusters on the 6N-$V_6$ monolayer (as will be shown later, this substrate gives metal dimers the highest activity for $CO_2$ reduction). As presented in Figure 1 and Figure S1, all metal dimers are embedded in the hole of the graphitic sheet, except that $Sc_2$ with a larger atomic size induces a noticeable buckling of 0.94 Å in the out-of-plane direction. Four N−metal bonds are formed with bond length of 1.95 ~ 2.09 Å, and the metal−metal bond length ranges from 1.96 Å to 2.79 Å (Table 2). The binding energy (defined by Equation S2 in Supplemental Information) between the metal dimer and the graphitic sheet is −4.29 ~ −10.28 eV, excluding the possibility of dissociation or aggregation of the metal dimer. The thermal stability of these carbon substrate anchored metal dimers was further assessed by *ab initio* molecular dynamics (AIMD) simulations, which manifest that they can sustain at least 800 K for 10 ps with small vertical displacement of metal atoms (< 0.2 Å) (see Figure S2 for details), suggesting superior thermal stability for practical uses.

A $CO_2$ molecule can favorably chemisorb on these dispersed metal dimers except $Cu_2$. The molecule is bended in the bidentate configuration with O−C−O angle of 124.90 ~ 141.96°. The C atom and one of the O atoms of $CO_2$ form two bonds with the underlying metal atoms; the C−O bond length is elongated to 1.21 ~ 1.36 Å, compared to 1.16 Å for a free $CO_2$ molecule. The adsorption energy (defined by Equation S3 in Supplemental Information) of $CO_2$ ranges from −0.82 eV to −3.40 eV. Overall speaking, stronger binding is provided by the metal element with fewer $d$ electrons. The trend of activity can be understood by the electronic density of states (DOS) shown in Figure 2a. Taking $Fe_2$@6N-$V_6$ as an example, hybridization between the $d$ orbitals of $Fe_2$ dimer and the $p$ orbitals of 6N-V$_6$ monolayer substrate is evident, with prominent electronic states at the vicinity of the Fermi level mainly contributed by the Fe atoms. Electron transfer of 0.73 $e$ occurs from Fe$_2$ to 6N-V$_6$ monolayer, which lifts the Fermi level of the hybrid system above the 2π* state of CO$_2$. As a result, Fe$_2$@6N-V$_6$ can favorably donate about 0.71 electrons to the antibonding orbital of CO$_2$, as manifested by the differential charge densities in Figure 3a. As depicted in Figure 2b, CO$_2$ adsorption energy generally follows a linear relationship with the $d$ band center of the supported metal dimers (relative to the Fermi level), as the metal dimer with a higher $d$ band center would provide stronger binding with CO$_2$ (Hammer et al., 2000).

In addition, we examined the capability of various dispersed 3$d$ transition metal dimers for activating two CO$_2$ molecules simultaneously, which is a prerequisite for C−C coupling to yield C$_2$ products. Several candidate systems including Sc$_2$, Ti$_2$, Cr$_2$, Mn$_2$ and Fe$_2$ dimers on the 6N-V$_6$ monolayer have adsorption energies of −3.71 ~ −0.48 eV for fixation of two CO$_2$ molecules (Figure 3 and Figure S3), while the other metal dimers are only able to bind one CO$_2$ molecule. Considering that Fe is an earth-abundant element and dispersed Fe atoms and dimers can be readily obtained in the experiment (Ye et al., 2019; Tian et al., 2018), thereafter we explored Fe$_2$ dimer on various nitrogenated 2D holey carbon materials as a representative of dual metal centers.

Figure 1 presents the structures of a Fe$_2$ dimer immobilized on several 2D carbon substrates. The dimer forms 4 ~ 6 bonds with the neighboring N or C atoms, having bond lengths of 1.91 ~ 2.21 Å for Fe−Fe and 1.87 ~ 2.00 Å for N−Fe (C−Fe) bonds, respectively, and the binding energies are −5.01 ~ −12.03 eV (Table 1). The Fe$_2$ dimer exhibits different buckling height in the out-of-plane direction (0.01 ~ 2.06 Å), and meanwhile induces some local vertical distortions on the carbon basal plane (0.09 ~ 0.35 Å). The dimer-substrate coupling strength depends on the size of the hole as well as the saturation degree of the edge atoms. For instance, binding strength between Fe$_2$ and 4N-V$_2$, 5N-V$_3$, and 6N-V$_4$(a) increases with both N content and hole size. The bonding interaction between Fe$_2$ and g-C$_3$N$_4$ or C$_2$N is relatively weak, due to the electronic saturation of these two semiconducting carbon nitride monolayers (as manifested by their large band gaps). In sharp contrast, $Fe_2$ is strongly anchored on the nitrogen-free $V_6$ defect that has six unsaturated carbon atoms on the hole edge, thereby leading to the largest binding energy of $-12.03$ eV.

All the supported $Fe_2$ dimers are able to chemisorb two $CO_2$ molecules with total adsorption energies of $-0.23 \sim -1.62$ eV (compared to $-0.11 \sim -1.58$ eV for adsorption of single $CO_2$ molecule), as revealed by Figure 3b. Both $CO_2$ molecules are bended with O−C−O angle of $141.00 \sim 152.13°$ and elongated C−O bond lengths of $1.17 \sim 1.29$ Å. The C atom in each $CO_2$ is bonded to the underlying Fe atom with Fe−C bond length of $1.93 \sim 2.12$ Å. Furthermore, we investigated the interaction between the dispersed $Fe_2$ dimers and the CO molecule, which is an important reaction intermediate in the $CO_2$ reduction process. Our calculations indicate strong binding of CO on the anchored $Fe_2$ dimers, with adsorption energies of $-2.94 \sim -4.04$ eV ($-1.94 \sim -2.70$ eV) for two (one) CO molecules. Consequently, desorption of CO from dual metal centers would be rather difficult, which allows further protonation of CO and thus provides the opportunity for successive C−C coupling.

The distinct binding capability of various supported $Fe_2$ dimers with gas molecules can be related to the electronic coupling between $Fe_2$ and the carbon substrate. As displayed in Figure 2c, the amount of charge transfer from $Fe_2$ to the substrate varies from $0.71$ $e$ to $0.97$ $e$. Generally speaking, less electron transfer leads to higher activity of the $Fe_2$ dimer for $CO_2$ and CO chemisorption, which is consistent with the trend of binding energies between $Fe_2$ and the carbon templates discussed before (Table 1). It is the N content, the degree of electronic saturation of the hole edge, and the bond configuration of $Fe_2$ in the hole, that jointly determine the coupling strength between the metal dimer and the carbon sheets. Therefore, the nitrogenated 2D holey carbon materials with diverse morphologies and controllable N contents can not only stabilize and disperse metal dimers, but also dictate the electronic structures and activity of the anchored metal dimers. This opens up a universal avenue for modulating the product selectivity for $CO_2$ reduction, as will be discussed in the following contents.

Note that graphitic N species are inevitably present in the experimentally synthesized N-doped carbon materials (Lin et al., 2014). To clarify their effect on the activity of the dispersed $Fe_2$ dimer, we investigated the $CO_2$ adsorption on $Fe_2$@6N-$V_6$ containing various numbers of graphitic N atoms at different distances from the 6N-$V_6$ hole (Figure S4). For all the considered systems, the $CO_2$ adsorption energies on the catalysts with and without substitutional N atoms on the graphene lattice differ by less than 0.16 eV, suggesting that existence of the graphitic N species has only minor impact on the catalytic properties of the $Fe_2$ dimer supported on pyridine holes of 2D carbon substrates.

Figure 4 shows the most efficient pathways for $CO_2$ reduction toward possible $C_1$ and $C_2$ products on the $Fe_2$ dimer immobilized on various nitrogenated carbon sheets, and the corresponding free energy diagrams of various model systems calculated by the computational hydrogen electrode (CHE) model (Peterson et al., 2010) are given by Figure 5, Figure S5 and Figure S6. We used point (.) to represent the co-adsorption of two carbon intermediates on the catalyst and strigula (−) to indicate the coupling between two carbon intermediates. The maximum Gibbs free energy of formation $\Delta G$ among all the reaction steps defines the rate-determine step (RDS) and is thus denoted as $\Delta G_{RDS}$. Overall speaking, formation of $C_2$ products first requires the activation of dual $CO_2$ molecules on the catalyst. By going through the carboxyl (COOH*) pathway, two CO* intermediates can be generated. Then, protonation of CO* leads to $C_1$ products like methanol ($CH_3OH$) and $CH_4$. Alternatively, it paves a way to the coupling between two neighboring carbon intermediates, which is energetically favorable and kinetically easy, and finally yields $C_2$ products ($C_2H_5OH$ and $C_2H_4$).

Specifically, formation of two CO* species on most of the considered $Fe_2$ dimers is uphill in the free energy profile, involving energy steps of 0.24 ~ 0.87 eV. Then, reduction of CO* gives rise to HCO* species, which is lower in energy by up to 1.13 eV than the other possible intermediates such as COH* (Figure S7). The CO* → HCO* conversion is endothermic with $\Delta G$ = 0.49 ~ 1.01 eV. Further protonation of HCO* leads to HCOH*, and then produces a CH* species by release of a $H_2O$ molecule. The C−C coupling reaction is most likely to occur between a CH* (or $CH_2$*) species and the neighboring CO*. Our nudged elastic band (NEB) calculations suggest that the CO−CH* coupling is exothermic and barrierless on all the considered $Fe_2$ dimers, except for $Fe_2@C_2N$ and $Fe_2@C_3N_4$ that involve a small kinetic barrier of about 0.22 eV (Table S1). According to previous theoretical studies (Goodpaster et al., 2016; Jiang et al., 2018), Cu(211) and (100) as the typical active surfaces for $CO_2$ reduction, favor dimerization of CO* or CO−HCO* coupling involving $\Delta G = -0.17 \sim 0.48$ eV. For the present $Fe_2$ dimers on nitrogenated carbon sheets, however, CO−CO* or CO−HCO* coupling has higher $\Delta G$ than the values of CO−CH* by 0.84 ~ 2.42 eV, and thus is unlikely to occur.

Following the C−C coupling, successive reduction of CO−CH* leads to CO−$CH_2$*, HCO−$CH_2$*, HCOH−$CH_2$*, HCOH−$CH_3$*, and finally yields $C_2H_5OH$. Alternatively, reduction of HCOH−$CH_2$* can give rise to CH−$CH_2$* with release of a $H_2O$ molecule, and further protonation of CH−$CH_2$* eventually produces $C_2H_4$. These elementary reactions involve relatively small steps of 0.15 ~ 0.73 eV in free energy profile, and thus would take place readily from the thermodynamic point of view. At the last step, desorption of $C_2H_5OH$* and $CH_2CH_2$* is endothermic by 0.11 ~ 0.59 eV and 0.23 ~ 1.76 eV, respectively. For most of the considered $Fe_2$ dimers, the rate-determine step for $C_2H_5OH$ production is the CO* → HCO* conversion, while release of $C_2H_4$ mainly suffers from the strong binding of $CH_2CH_2$* on the catalyst.

On the other hand, formation of $C_1$ products is also possible on the dispersed $Fe_2$ dimers. As discussed above, HCOH* can be reduced to CH*, followed by the CO−CH* coupling. Alternatively, HCOH* may be protonated to $H_2COH$*. Then, reduction of $H_2COH$* yields $CH_3OH$, or produces $CH_2$* with release of a $H_2O$ molecule followed by the generation of $CH_3$* and $CH_4$. For $Fe_2@4N-V_2$, $Fe_2@6N-V_6$ and $Fe_2@g-C_3N_4$, the CO* → HCO* conversion is the rate-determine step for both $C_1$ products. For $Fe_2@6N-V_4(b)$ and $Fe_2@5N-V_3$, formation of $CH_3OH$ from $H_2COH$* protonation requires $\Delta G_{RDS} = 1.45$ and 0.98 eV, respectively. In particular, $Fe_2@C_2N$ encounters $\Delta G_{RDS} = 0.94$ eV and a kinetic barrier of 0.77 eV during the reaction of HCOH* → $H_2COH$* for both $C_1$ products, whereas the competing step of CO−HCOH* →

CO−CH* + H$_2$O has much reduced $\Delta G$ = −0.30 eV and a lower kinetic barrier of 0.42 eV (Figure 5a, c). This would lead to high selectivity for C$_2$ products on Fe$_2$@C$_2$N.

Figure 6(a) plots $\Delta G_{RDS}$ values for various C$_1$ and C$_2$ products from CO$_2$ reduction on the anchored Fe$_2$ dimers. Among the four products, C$_2$H$_5$OH exhibits the lowest $\Delta G_{RDS}$ = 0.57 ~ 1.01 eV, and the highest activity is achieved by Fe$_2$@6N-V$_6$ owing to its moderate adsorption strength with the reaction intermediates (indicated by the dashed blue line in Figure 6a). Formation of C$_2$H$_4$ is less favorable with $\Delta G_{RDS}$ = 0.58 ~ 1.76 eV due to the strong binding of CH$_2$CH$_2$* on the Fe$_2$ dimers. Fe$_2$@6N-V$_4$(a), Fe$_2$@4N-V$_2$, Fe$_2$@6N-V$_6$ and Fe$_2$@g-C$_3$N$_4$ exhibit similar selectivity for C$_2$H$_5$OH, CH$_3$OH, and CH$_4$, while Fe$_2$@5N-V$_3$ favors both C$_2$H$_5$OH and CH$_4$ products. Remarkable selectivity for C$_2$H$_5$OH is obtained for Fe$_2$@C$_2$N and Fe$_2$@6N-V$_4$(b) with $\Delta G_{RDS}$ = 0.70 and 0.59 eV, respectively, notably lower than $\Delta G_{RDS}$ values for the other products (above 0.94 and 0.85 eV, respectively). Hence, these supported Fe$_2$ dimers have competitive activity but distinct selectivity with regard to the conventional Cu based catalysts. It is known that Cu crystals mainly produce CO under low electrode potentials, while CH$_4$ and C$_2$H$_4$ are the main products at sufficiently high electrode potentials (about −1.0 V vs. RHE in experiment) (Dai et al., 2017; Mistry et al., 2016). Previous calculations revealed that Cu(211) surface encounters $\Delta G_{RDS}$ = 0.74 eV for CH$_4$ and C$_2$H$_4$, while formation of CO is much more favorable with $\Delta G_{RDS}$ = 0.41 eV due to the relatively weak adsorption of CO on the Cu surface (adsorption energy $\Delta E$ = −1.01 eV) (Peterson et al., 2010). Differently, release of CO is prohibited on the present Fe$_2$ dimers that have strong adsorption energy of $\Delta E$ = −2.94 ~ −4.04 eV with CO molecule.

The unique geometry and favorable adsorption properties of the Fe$_2$ dimers immobilized on carbon substrates bring about inimitable advantages for their catalytic behavior. First, CO as an inevitable and even dominant product of CO$_2$ reduction on many metal catalysts, severely limits the formation of higher-energy-density products (Zhu et al., 2014; Sarfraz et al., 2016; Peng et al., 2018); but it would be largely suppressed on the anchored Fe$_2$ dimers. Second, the adjacent dual metal centers and their strong binding with CO pave an efficient pathway for C−C coupling reaction; in contrast, C−C coupling only occurs on metal surfaces with homogenously distributed reaction sites when the coverage of CO is sufficiently high (Morales-Guio et al., 2018; Huang et al., 2017). Third, the difficult desorption of $C_2H_4$ from the $Fe_2$ dimers may result in superior selectivity for $C_2H_5OH$, which is a clean liquid fuel with high heating value. For most of the Cu based catalysts, however, the yield of $C_2H_5OH$ is quite low compared to $C_2H_4$ (Liang et al., 2018).

At last, we assess the activity of these supported $Fe_2$ dimers for HER, which is a competing reaction against $CO_2$ reduction and highly affects the efficiency of $CO_2$ conversion (Zhu et al., 2016; Cui et al., 2017). Figure 6(b) plots the competition between adsorption of H* species and $CO_2$ molecule on the $Fe_2$ dimers. The H* adsorption energy ranges from −1.52 eV to −0.28 eV. For $Fe_2@5N-V_3$, $Fe_2@6N-V_6$, $Fe_2@C_2N$ and $Fe_2@C_3N_4$, the adsorption strength of H* species is notably weaker than that of $CO_2$ molecule by 0.09 ~ 0.67 eV, implying that $CO_2$ reduction would prevail over HER on these catalysts with either high activity or superior selectivity. For $Fe_2@4N-V_2$, $Fe_2@6N-V_4(a)$ and $Fe_2@6N-V_4(b)$, the H* adsorption strength is stronger than that of $CO_2$, which may suppress the $CO_2$ reduction. In short, our results confirm that the $Fe_2$ dimers immobilized on various nitrogenated carbon materials universally possess outstanding activity for $C_2H_5OH$ synthesis from $CO_2$ reduction, and desired selectivity can be further achieved by selecting proper carbon substrates.

**CONCLUSION**

In summary, we exploited dispersed 3*d* transition metal dimers for $CO_2$ reduction to selectively produce liquid fuels. Comprehensive first-principles calculations show that nitrogenated holey carbon materials not only serve as templates to stabilize small metal clusters, but also dictate their electronic structures. Specifically, controlling the metal-substrate coupling strength allows effective modulation of both activity and product selectivity. As a consequence, the spatially confined dual reaction centers within the carbon matrix exhibit the following advantageous catalytic behavior: (1)

simultaneous fixation of two $CO_2$ molecules, (2) prohibition of CO desorption, (3) exclusive pathway for C−C coupling, (4) high activity for $C_2H_5OH$ production irrelevant to the type of substrate. A number of immobilized dimer systems with outstanding activity and selectivity have been obtained. In particular, a $Fe_2$ dimer embedded in the $C_2N$ monolayer exhibits remarkable selectivity for $C_2H_5OH$ against the other $C_1$ and $C_2$ products as well as HER. These theoretical findings provide vital guidance for atomically precise design of the dispersed metal clusters for converting the greenhouse gas to high-energy fuels and high-value chemicals.

**Limitations of the Study**

This study systematically exploited $3d$ transition metal dimers anchored on nitrogenated holey carbon monolayers for selective reduction of $CO_2$ to liquid fuels, and screened suitable metal elements and carbon templates with high selectivity for ethanol. However, experimental realization of such superior subnano catalysts relies on the preparation of metal clusters with specific size supported on some given substrates, which may be challenging and requires the development of advanced synthesis methods.

**METHODS**

All methods can be found in the accompanying Transparent Methods supplemental file.

**Acknowledgements**

This work was financially supported by the National Natural Science Foundation of China (11974068, 91961204). The authors acknowledge the computer resources provided by the Supercomputing Center of Dalian University of Technology.

**Author Contributions**

S. Zhou conceived the idea; Y. Zhao carried out the calculation; S. Zhou and J. Zhao supervise the research. All authors wrote the paper.

**Declaration of Interests**

The authors declare no competing interests.

**Table 1.** Formation energy ($E_{form}$) of various nitrogenated 2D holey carbon materials, binding energy ($E_b$) of a $Fe_2$ dimer on the carbon sheet, bond length ($d$) of Fe−Fe and N−Fe/C−Fe bonds, Mulliken charge transfer (CT) from $Fe_2$ to the carbon sheet, adsorption energy of a $CO_2$ molecule ($\Delta E_{CO2*}$) on the supported $Fe_2$ dimer.

| Substrate | $E_{form}$ (eV/Å) | $E_b$ (eV) | $d$ (Å) Fe−Fe | $d$ (Å) N−Fe | CT ($e$) | $\Delta E_{CO2*}$ (eV) |
|---|---|---|---|---|---|---|
| 4N-V$_2$ | 0.17 | −5.01 | 2.09 | 1.98 | 0.71 | −1.02 |
| 5N-V$_3$ | 0.19 | −7.33 | 1.91 | 1.87 | 0.97 | −1.15 |
| 6N-V$_4$(a) | 0.20 | −9.47 | 2.21 | 1.94 | 0.96 | −0.11 |
| 6N-V$_4$(b) | 0.21 | −7.48 | 2.14 | 1.98 | 0.81 | −1.15 |
| 6N-V$_6$ | 0.16 | −6.02 | 1.96 | 2.00 | 0.76 | −1.12 |
| C$_2$N | -- | −5.80 | 2.01 | 1.97 | 0.74 | −0.64 |
| g-C$_3$N$_4$ | -- | −5.09 | 1.98 | 1.99 | 0.72 | −1.58 |
| V$_6$ | 0.67 | −12.03 | 2.21 | 1.94 | 0.87 | −0.30 |

**Table 2.** Binding energy ($E_b$) of various 3$d$ transition metal dimers anchored on the 6N-V$_6$ monolayer, bond lengths ($d$) of metal dimer (M−M) and N−metal (N−M), adsorption energy of single and dual CO$_2$ molecules on the supported metal dimers ($\Delta E$), and the $d$ band center ($\varepsilon_d$) of the supported metal dimers (Hammer et al., 2000).

| Metal dimer | $E_b$ (eV) | $d$ (Å) | | $\Delta E$ (eV) | | $\varepsilon_d$ (eV) |
|---|---|---|---|---|---|---|
| | | M−M | N−M | CO$_2$ | 2CO$_2$ | |
| Sc$_2$ | −10.28 | 2.79 | 2.09 | −3.40 | −3.71 | 1.16 |
| Ti$_2$ | −8.50 | 2.17 | 1.99 | −2.85 | −3.31 | 0.62 |
| V$_2$ | −8.80 | 2.14 | 1.96 | −2.26 | -- | 0.41 |
| Cr$_2$ | −4.99 | 2.16 | 1.97 | −1.44 | −0.76 | 0.07 |
| Mn$_2$ | −6.52 | 2.04 | 2.01 | −1.05 | −0.48 | −0.42 |
| Fe$_2$ | −6.02 | 1.96 | 2.00 | −1.12 | −0.50 | −1.00 |
| Co$_2$ | −5.71 | 2.10 | 1.95 | −1.20 | -- | −1.09 |
| Ni$_2$ | −5.93 | 2.17 | 2.00 | −0.82 | -- | −1.12 |
| Cu$_2$ | −4.29 | 2.35 | 1.96 | −0.35 | -- | −2.08 |

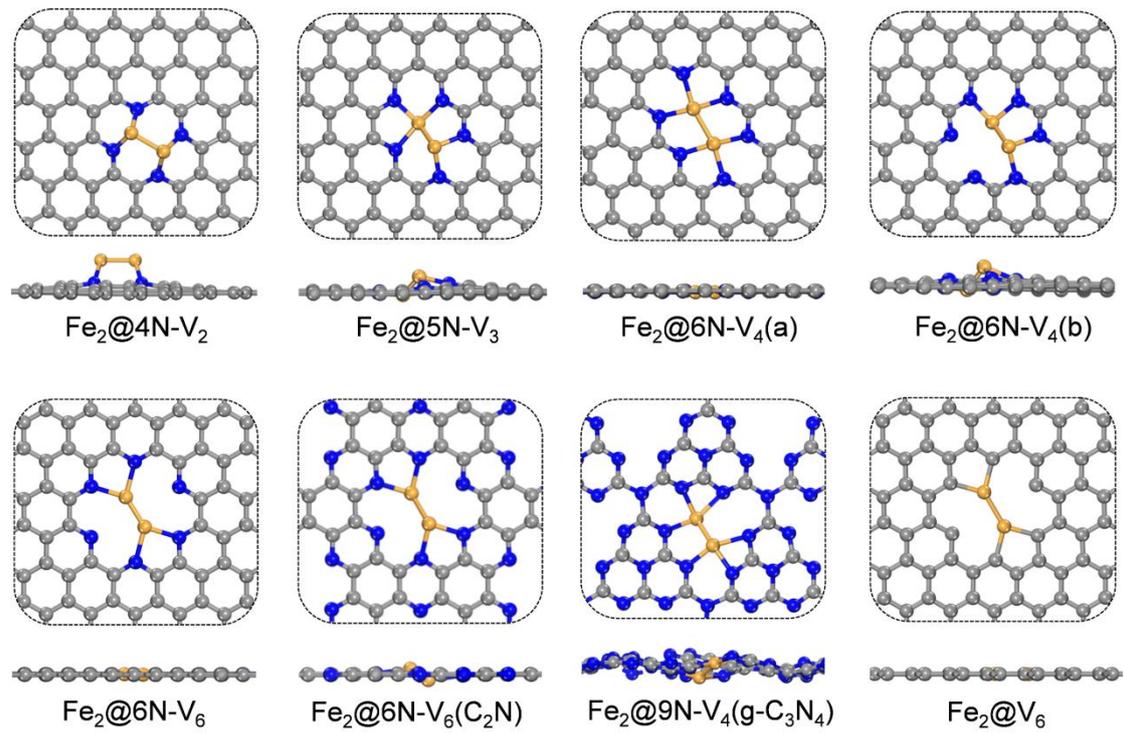

**Figure 1.** Atomic structures of a Fe$_2$ dimer anchored on various nitrogenated holey carbon monolayers (top panel: top view; bottom panel: side view). The C, N and Fe atoms are shown in grey, blue and orange colors, respectively.

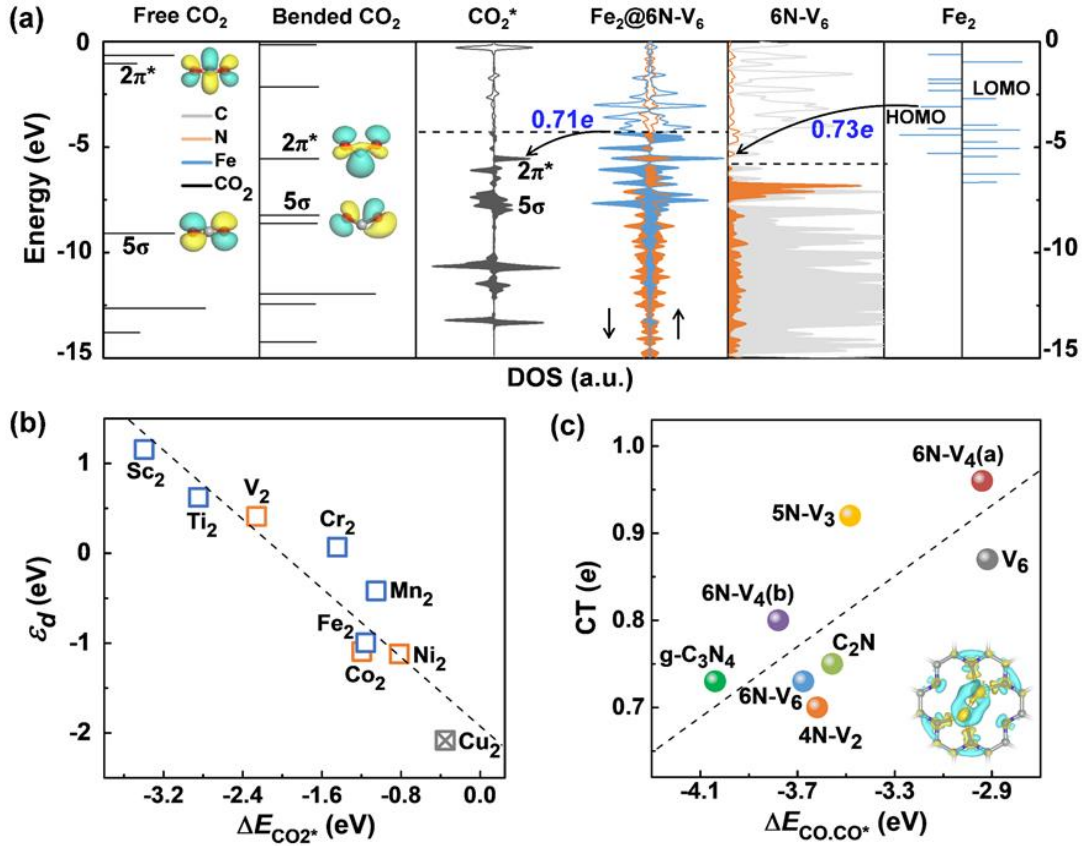

**Figure 2.** (a) From left to right: molecular orbital levels or local density of states (DOS) of a free and a bended (with C−O−C angle of 130°) $CO_2$ molecule in vacuum, an adsorbed $CO_2$ molecule on $Fe_2$@$6N$-$V_6$, an individual $6N$-$V_6$ monolayer and a $Fe_2$ dimer. The insets display the HOMO and LUMO charge densities of $CO_2$. The energy is relative to the vacuum. The dashed line shows the Fermi level, with the occupied states shadowed. (b) The $d$ band center ($\varepsilon_d$) of various supported $3d$ transition metal dimers as a function of the adsorption energy of single $CO_2$ molecule. The blue/orange/grey symbols denote that two/one/none $CO_2$ molecule can be chemisorbed on the metal dimer. The dashed line is a linear fit of the data points. (c) Charge transfer (CT) from the $Fe_2$ dimer to various nitrogenated carbon holey monolayer as a function of the adsorption energy of dual CO molecules. The dashed line is a linear fit of the data points. The insert shows the differential charge density of $Fe_2$@$6N$-$V_6$. The yellow and cyan colors represent the electron accumulation and depletion regions, respectively, with an isosurface value of 0.005 $e$/Å$^3$.

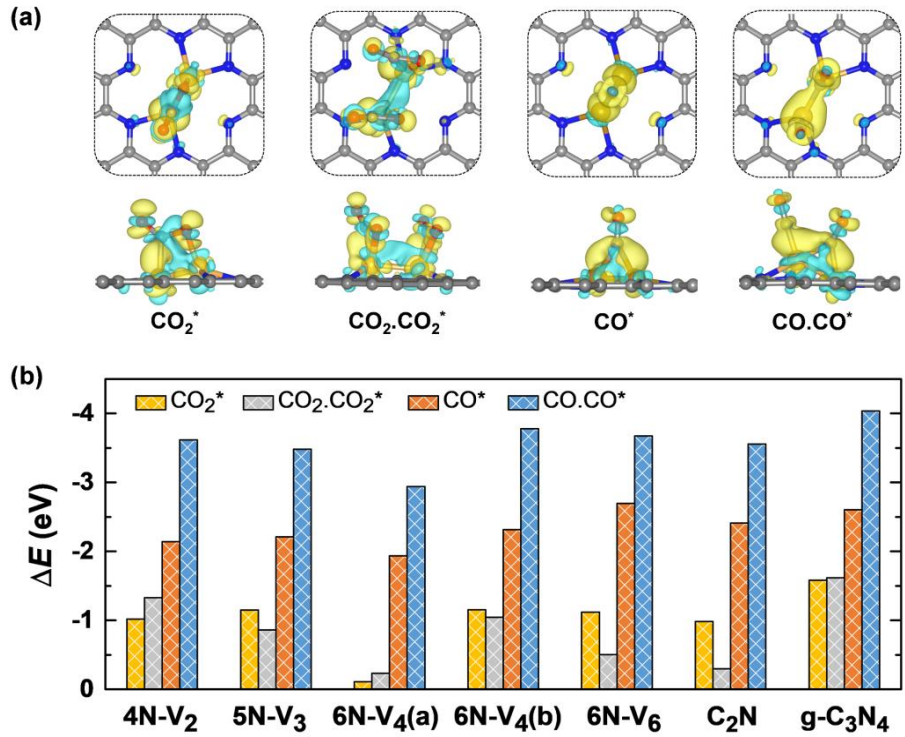

**Figure 3.** (a) From left to right: differential charge densities of single and dual $CO_2$ molecules, single and dual CO molecules adsorbed on Fe$_2$@6N-V$_6$. The yellow and cyan colors represent the electron accumulation and depletion regions, respectively, with an isosurface value of 0.005 $e$/Å$^3$. (b) Adsorption energies of single and dual $CO_2$ and CO molecules on the Fe$_2$ dimer anchored on various nitrogenated holey carbon monolayers. The C, N, O and Fe atoms are shown in grey, blue, red and orange colors, respectively.

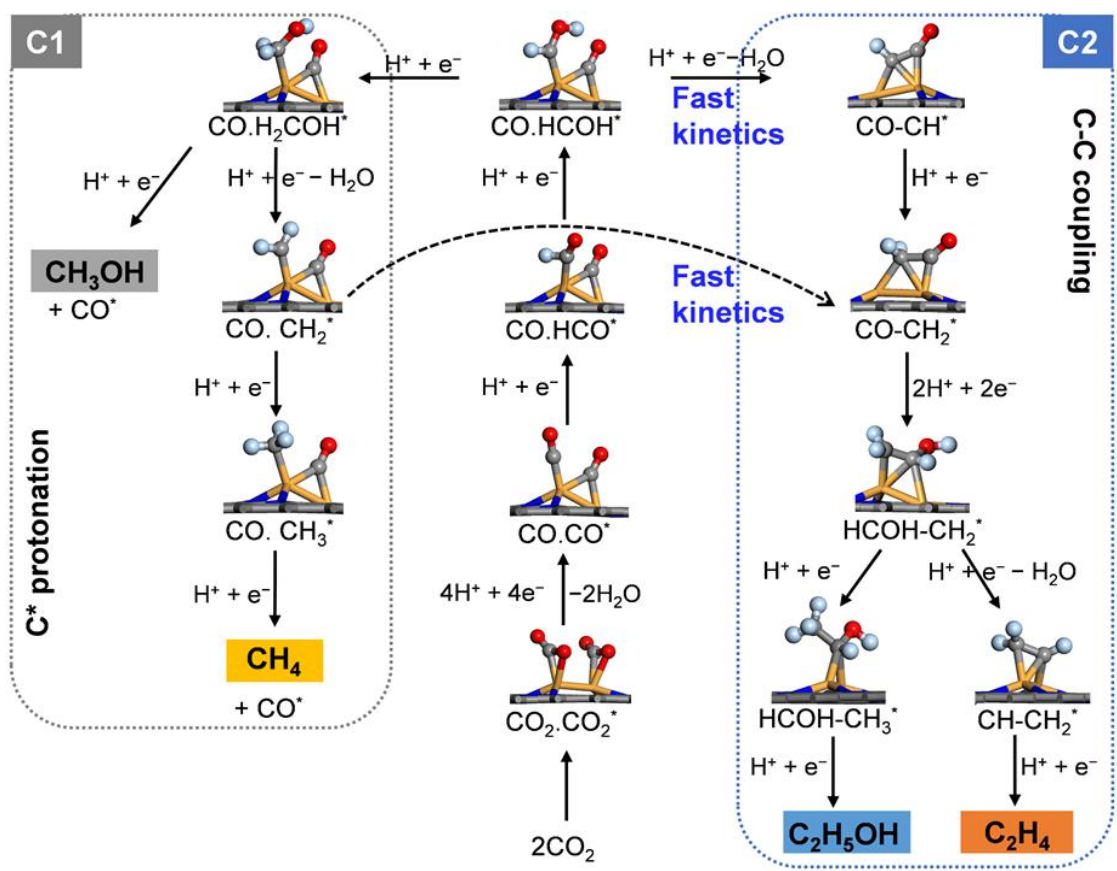

**Figure 4.** The $CO_2$ reduction pathways to various $C_1$ and $C_2$ products on the supported $Fe_2$ dimer. The H, C, N, O and Fe atoms are shown in light blue, grey, blue, red and orange colors, respectively.

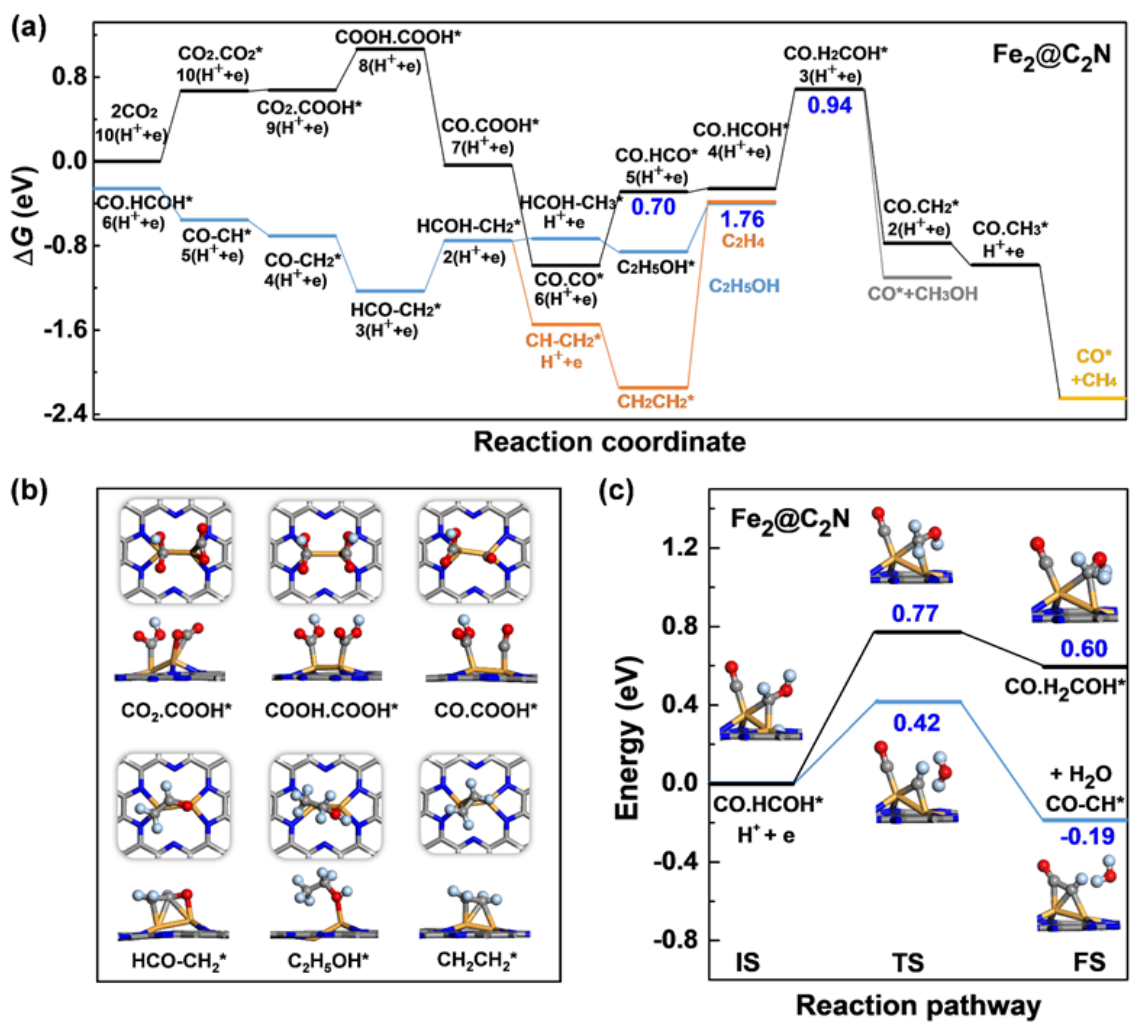

**Figure 5.** (a) Free energy diagram of $CO_2$ reduction to various $C_1$ and $C_2$ products (indicated by different colors) on $Fe_2@C_2N$. The blue numbers, from left to right, give the Gibbs free energy of formation for the rate-determine step of $C_2H_5OH$, $C_2H_4$ and $CH_3OH/CH_4$. The local structures of selected reaction intermediates are presented in (b). The H, C, N, O and Fe atoms are shown in light blue, grey, blue, red and orange colors, respectively. (c) Competing reactions of CO.HCOH to form $C_1$ and $C_2$ intermediates on $Fe_2@C_2N$. The insets display the structures of initial state (IS), transition state (TS) and final state (FS). The blue numbers give the kinetic barriers (middle) and heat of reaction (right).

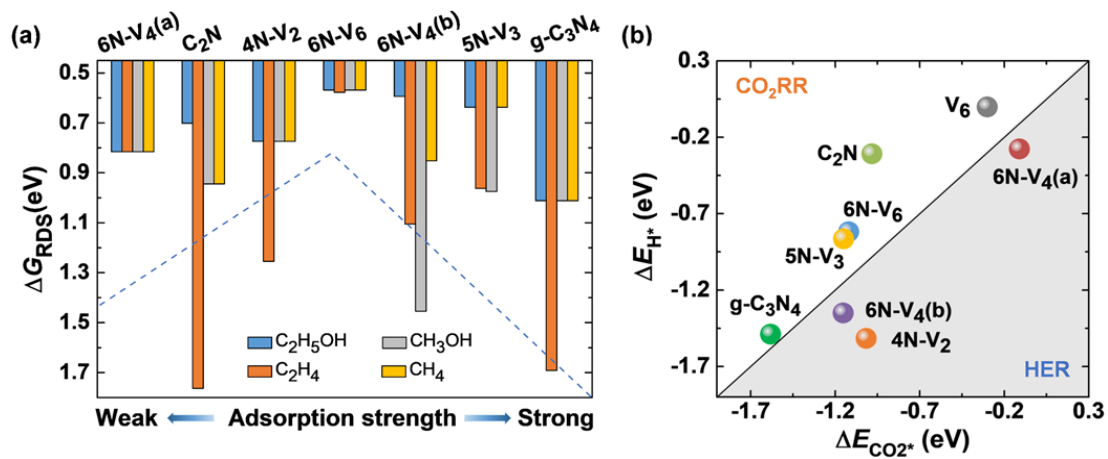

**Figure 6.** (a) Gibbs free energy of formation for the rate-determine step ($\Delta G_{RDS}$) for various $C_1$ and $C_2$ products from $CO_2$ reduction, and (b) competition between adsorption of a $CO_2$ molecule and a H* species on the $Fe_2$ dimer anchored on various nitrogenated holey carbon monolayers.